\documentclass[aps,preprint,eqsecnum]{revtex4}
\usepackage{epsfig}
\usepackage{graphicx}
\usepackage{psfrag}
\newcommand{\be}{\begin{eqnarray}}
\newcommand{\ee}{\end{eqnarray}}

\begin{document}
\vspace*{1cm}
\title
{Generalized Parton Distributions in the Impact Parameter Space with Non-zero
Skewedness}
\author{\bf D. Chakrabarti}
\email{dipankar@phys.ufl.edu}
\affiliation{Department of Physics, University of Florida, Gainesville,
FL-32611-8440, USA}
\author{\bf A. Mukherjee}\email{asmita@lorentz.leidenuniv.nl}
\affiliation{ Instituut-Lorentz, University of Leiden, 2300 RA Leiden,
The Netherlands}

\date{\today\\[2cm]}

\begin{abstract}
We investigate the generalized parton distributions (GPDs) with non-zero
$\xi$ and $\Delta^\perp$ for a
relativistic spin-$1/2$ composite system, namely for an electron dressed
with a photon,  in light-front framework 
by expressing them in terms of overlaps of
light-cone wave functions. The wave function provides a template for the
quark spin-one diquark structure of the valence light cone wave function of
the proton. We verify the inequalities among
the GPDs with different helicities and show the qualitative behaviour of the 
fermion and gauge boson GPDs in the impact parameter space.
\end{abstract}
\maketitle

\section{Introduction}
Generalized parton distributions (GPDs) have attracted a considerable amount
of theoretical and experimental attention recently. An interesting physical
interpretation of GPDs has been obtained in \cite{bur1,bur2} by taking their
Fourier transform with respect to the transverse momentum transfer. 
When the longitudinal momentum transfer $\xi=0$, this gives the distribution
of partons in the nucleon in the transverse plane. They are called impact
parameter dependent parton distributions (ipdpdfs) $q(x,b^\perp)$. In fact
they obey certain positivity constraints which justify their physical
interpretation as probability densities. This interpretation holds in the
infinite momentum frame (even the forward pdfs have a probabilistic
interpretation only in this frame)  and there is no relativistic correction to this
identification because in light-front formalism, as well as in the infinite momentum 
frame, the transverse boosts act like non-relativistic Galilean boosts. It is to 
be remembered that the
GPDs, being off-forward matrix elements of light-front bilocal currents do
not have a probabilistic interpretation, rather they have interpretation as
probability amplitudes. $q(x,b^\perp)$ is defined in a proton state with      
a sharp plus momentum $p^+$ and localized in the transverse plane such that
the transverse center of momentum $R^\perp=0$ (normally, one should work with
a wave packet state which is very localized in transverse position space, in
order to avoid the state to be normalized to a delta function
\cite{bur2,diehl}). 
$q(x, b^\perp)$ gives simultaneous information about the longitudinal
momentum fraction $x$ and the transverse distance $b$ of the parton from the
center of the proton and thus gives a new insight to the internal structure
of the proton. The impact parameter space representation has also been
extended to the spin-dependent GPDs \cite{bur1} and chiral odd ones
\cite{chiral}.

GPDs $H_q(x,0,t)$ have been investigated in the impact parameter space in
several approaches, for example in the transverse lattice formalism for the
pion \cite{dalley}, in a two component (spectator) model \cite{liuti} for
the nucleon, in the chiral quark model for the pion \cite{arriola} and using
a power law wave function for the pion \cite{musatov}. 
The spin-flip GPD $E_q$
has not been addressed in these. The connection of $E_q$ in the impact
parameter space and the Siver's effect has been shown in \cite{siv} within
the framework of the scalar diquark model of the proton. In a previous work \cite{imp1}, we have calculated both
$H(x,0,t)$ and $E(x,0,t)$ in the impact parameter space for a spin-${1/2}$
composite relativistic system, namely for an electron dressed with a photon
in QED. The state can be expanded in Fock space in terms of light-cone
wave functions. The GPDs are expressed as overlaps of light-cone wave
functions \cite{overlap}. The  wave functions in this case can be obtained
from perturbation theory, and thus their correlations are known at a certain
order in the coupling constant. Their general form provides a
template for the effective quark spin-one diquark structure of the valence
light-cone wave
function of the proton \cite{brod1}. Such a model is self consistent and 
has been used to investigate the helicity structure of a composite relativistic system
\cite{brod1}. An interesting advantage is that the 
two-body Fock 
component contains a gauge boson as one of its constituents and so it is 
possible to
investigate the gauge boson GPDs $H_g$ and $E_g$. Studies of the deep
inelastic scattering structure functions in this approach and for a dressed
quark state also yield interesting results \cite{hari,hari1}.   
     
So far we have discussed GPDs in impact parameter space for $\xi=0$. 
However, deeply virtual Compton scattering experiments probe GPDs at nonzero
$\xi$. In this case, a Fourier transform with respect to the transverse 
momentum transfer $\Delta^\perp$ is not enough to diagonalize the GPDs and
thus giving a density interpretation. As the longitudinal momentum in the
final state is different from that in the initial state, the resulting
matrix element would still be off-diagonal. Recently, certain reduced Wigner
distributions, when integrated over the transverse momenta of the partons
are shown to be the Fourier transforms of GPDs \cite{wigner} and they
can be interpreted as the 3D density in the rest frame of the proton for the
quarks with light cone momentum fraction $x$. In fact, integration over the
$z$ coordinate  relates them  to
the ipdpdfs with $\xi=0$. In \cite{diehl} it has been shown that for nonzero
$\xi$, the Fourier transform of the GPDs with respect to $\Delta^\perp$
probes partons at transverse position $b^\perp$, with the initial and final
protons localized around $0^\perp$ but shifted from each other by an amount
of order $\xi b^\perp$. At the same time, the longitudinal momentum of the
protons are specified. This difference of the transverse position of the
protons depends on $\xi$ but not on $x$ and thus this information should be
present in the scattering amplitudes measurable in experiments where the
GPDs enter through a convolution in $x$.  This aspect makes it interesting
to investigate the GPDs in the impact parameter space for nonzero $\xi$.
Here also, a
useful approach is based on the overlap representation of GPDs in terms of
light-cone wave functions \cite{overlap}. The overlap representation can
also be formulated directly in the impact parameter space, in terms of
overlaps of light-cone wave functions $\psi(x, b^\perp)$, which are the Fourier
transforms of the wave functions with definite transverse momenta $\psi(x,
k^\perp)$. 

Here, we calculate the GPDs $H_{q,g}(x, \xi,t)$ and $E_{q,g}(x, \xi, t)$ for
an effective spin-${1/2}$ system of an electron dressed with a photon in 
QED and we investigate them in the
impact parameter space. The plan of the paper is as follows.   
The definitions of the fermion and gauge boson GPDs are given in section II.
 The fermion and gauge boson GPDs 
are calculated respectively in section III and IV for a dressed 
electron state. The GPDs are expressed in the impact
parameter space in section  V. The issue of certain  
inequalities among the GPDs in the impact parameter space is addressed in 
section VI. The summary and discussions are given in section VII.  

\section{Generalized Parton Distributions}
The GPDs are defined in terms of off-forward matrix elements of light-front
bilocal currents. In the light-front gauge $A^+=0$ we have, 
\be
F^{+q}_{\lambda' \lambda}&=&\int {dy^-\over {8 \pi}} e^{{i\over 2} x
{\bar P}^+{y^-}}\langle P'
\lambda' \mid \bar \psi (-{ y^-\over 2}) \gamma^+ \psi ({ y^-\over 2})
\mid P \lambda \rangle \nonumber\\&&={1\over { 2{\bar P}^+}}
{\bar U}_{\lambda'}(P') 
\Big [ H_q(x,\xi,t) \gamma^+ + E_q(x,\xi,t) {i\over {2M}}
\sigma^{+ \alpha} \Delta_\alpha \Big ] U_\lambda(P) +.....
\label{gpdq}
\ee
\be
F^{+g}_{\lambda' \lambda}&=&{1\over {8 \pi x {\bar P}^+}}\int dy^- 
e^{{i\over 2}{\bar P}^+ { y^-} x}\langle P'
\lambda' \mid F^{+ \alpha}(-{ y^-\over 2}) {F^+}_\alpha ({ y^-\over 2})
\mid P \lambda \rangle \nonumber\\&&={1\over { 2{\bar P}^+}}
{\bar U}_{\lambda'}(P') 
\Big [ H_g(x,\xi,t) \gamma^+ + E_g(x,\xi,t) {i\over {2M}}
\sigma^{+ \alpha} \Delta_\alpha \Big ] U_\lambda(P) +.....
\label{gpdg}
\ee 
where the ellipses indicate higher twist terms. The momenta of the initial
(final) state is $P (P')$ and helicity $\lambda (\lambda')$.
$U_\lambda(P)$ is the light-front spinor for the proton. 
The momentum transfer is given by $\Delta^\mu=P'^\mu-P^\mu$,
skewedness $\xi=-{\Delta^+\over {2 {\bar P}^+}}$. The average momentum of
the initial and final state proton is ${\bar P}^\mu={P^\mu+P'^\mu\over 2}$.
We take the frame where ${\bar P}^\perp=0$. Without any loss of generality, 
we take $\xi >0$. $t$ is the invariant momentum transfer in the process,
$t=\Delta^2$. For simplicity we suppress the flavor indices. 
Following \cite{diehl} we define
\be
D^\perp={P'^\perp\over 1-\xi}-{P^\perp\over 1+\xi}={\Delta^\perp\over
1-\xi^2}. 
\ee
$H_{q,g}$ and $E_{q,g}$ are the twist two fermion and gauge boson GPDs. 
Using the light-cone spinors
\cite{ped} we get
\be
F^{+q,g}_{++} = F^{+q,g}_{--} = {\sqrt {1-\xi^2}} H_{q,g}(x,\xi,t)
-{\xi^2 \over {\sqrt {1-\xi^2}}}E_{q,g}(x,\xi,t).
\label{he}
\ee
\be
F^{+ q,g}_{+-}=F^{+ q,g}_{-+}={-\Delta^1+i\Delta^2\over {2 M \sqrt
{1-\xi^2}}}E_{q,g}(x,\xi,t).
\label{heflip}
\ee
Note that $E_{q,g}$ appear both in helicity-flip and helicity non-flip
parts. For $\xi=0$, $H_{q,g}$ correspond to nucleon helicity non-flip and
$E_{q,g}$ correspond to the helicity flip part. 

\section{Fermion GPDs}
We take the state 
$ \mid P, \sigma \rangle$ to be a dressed electron
consisting of bare states of an electron and an electron plus
a photon :
\begin{eqnarray}
\mid P, \sigma \rangle && = \phi_1 b^\dagger(P,\sigma) \mid 0 \rangle
\nonumber \\  
&& + \sum_{\sigma_1,\lambda_2} \int 
{dk_1^+ d^2k_1^\perp \over \sqrt{2 (2 \pi)^3 k_1^+}}
\int 
{dk_2^+ d^2k_2^\perp \over \sqrt{2 (2 \pi)^3 k_2^+}}   
\sqrt{2 (2 \pi)^3 P^+} \delta^3(P-k_1-k_2) \nonumber \\
&& ~~~~~\phi_2(P,\sigma \mid k_1, \sigma_1; k_2 , \lambda_2) b^\dagger(k_1,
\sigma_1) a^\dagger(k_2, \lambda_2) \mid 0 \rangle.
\label{eq2}    
\end{eqnarray} 

Here $a^\dagger$ and $b^\dagger$ are bare photon and electron
creation operators respectively and $\phi_1$ and $\phi_2$ are the
multiparton wave functions. They are the probability amplitudes to find one
bare electron and one electron plus photon inside the dressed electron state
respectively.

We introduce Jacobi momenta $x_i$,${q_i}^\perp$ such that $\sum_i x_i=1$ and
$\sum_i {q_i}^\perp=0$.  They are defined as
\be
x_i={k_i^+\over P^+}, ~~~~~~q_i^\perp=k_i^\perp-x_i P^\perp.
\ee
Also, we introduce the wave functions,  
\be
\psi_1=\phi_1, ~~~~~~~~~~~\psi_2(x_i,q_i^\perp)= {\sqrt {P^+}} \phi_2
(k_i^+,{k_i}^\perp);
\ee
which are independent of the total transverse momentum $P^\perp$ of the
state and are boost invariant.
The state is normalized as,
\be
\langle P',\lambda'\mid P,\lambda \rangle = 2(2\pi)^3
P^+\delta_{\lambda,\lambda'} \delta(P^+-{P'}^+)\delta^2(P^\perp-P'^\perp).
\label{norm}
\ee
The two particle wave function depends on the helicities of the electron and
photon. Using the eigenvalue equation for the light-cone Hamiltonian, this
can be written as \cite{hari},
\be
\psi^\sigma_{2\sigma_1,\lambda}(x,q^\perp)&=& -{x(1-x)\over
(q^\perp)^2+m^2 (1-x)^2}
{1\over {\sqrt {(1-x)}}} {e\over
{\sqrt {2(2\pi)^3}}} \chi^\dagger_{\sigma_1}\Big [ 2 {q^\perp\over
{1-x}}+{{\tilde \sigma^\perp}\cdot q^\perp\over x} {\tilde \sigma^\perp}
\nonumber\\&&~~~~~~~~~~~~~~~~~~
-i m{\tilde \sigma}^\perp {(1-x)\over x}\Big ]\chi_\sigma
\epsilon^{\perp *}_\lambda \psi_1.
\label{psi2}
\ee
$m$ is the bare mass of the electron, $\tilde \sigma^2 = -\sigma^1$ and
$\tilde \sigma^1= \sigma^2$. 
$\psi_1$ actually gives the normalization of the state \cite{hari}:
\be
{\mid \psi_1 \mid}^2=1-{\alpha\over {2 \pi}} 
\int_\epsilon^{1-\epsilon} dx \Big [{{1+x^2}\over {1-x}}log
{\Lambda^2\over m^2 (1-x)^2}-{1+x^2\over 1-x}+(1-x) \Big ],
\label{c5nq}
\ee
within order $\alpha$.  Here $\epsilon$ is a small cutoff on $x$, the longitudinal momentum 
fraction carried by the fermion. We have
taken the cutoff of the transverse momenta to be $\Lambda^2$ \cite{hari}. 
This gives the large scale of the process. The above expression is derived using Eqs (\ref{norm}), (\ref{eq2}) and
(\ref{psi2}).

The helicity non-flip part of the matrix element $
F^{+q}_{++} $ gives information about both $H_q$ and $E_q$, as can be seen
from (\ref{he}). In terms of the wave function this can be written as,
\be
F^{+q}_{++} = {\mid \psi_1 \mid }^2 \delta (1-x)+\int d^2q^\perp \psi^*_{2+}({x-\xi\over 1-\xi},
q^\perp+(1-x) D^\perp) \psi_{2+}({x+\xi\over 1+\xi}, q^\perp).
\label{qnonflip}
\ee
We restrict ourselves to the DGLAP region $1>x>\xi$. As we are considering
no antiparticles $0<x<1$ in our case. It is known that in the ERBL region,
$-\xi<x<\xi$ the GPDs are expressed as off-diagonal overlaps of light-cone
wave functions involving higher Fock components \cite{overlap}.

The normalization of the state, ${\mid \psi_1 \mid }^2$ given by eq.
(\ref{c5nq}) gives another $O(\alpha)$ contribution. The $q^\perp$ integral
in the above expression is divergent and can be performed using the same
cutoffs as discussed above. We get, using Eq. (\ref{psi2}),
\be
F^{+q}_{++} &=& {\mid \psi_1 \mid }^2 \delta (1-x) + {e^2\over 2 (2 \pi)^3} 
{\mid \psi_1 \mid }^2  {1\over \sqrt {(1-x_1) (1-x_2)} } \Big \{ 2 (1+ x_1
x_2) \nonumber\\&&~~~\Big [ \pi log[{\Lambda^2\over {(1-x)^2 D_\perp^2 +m^2 (1-x_1)^2}}]-m^2
(1-x_2)^2 I \Big ]+ (1+x_1 x_2) \Big [ \pi log[{\Lambda^2\over m^2 (1-x_2)^2}] 
\nonumber\\&&~~~-\pi log[{\Lambda^2\over (1-x)^2 D_\perp^2 +m^2 (1-x_1)^2}] -\Big ( (1-x)^2
D_\perp^2 +m^2 [ (1-x_1)^2-(1-x_2)^2 ] \Big ) I \Big ]
\nonumber\\&&~~~~~~~~~+ 2 m^2 (1-x_1)^2
(1-x_2)^2 I \Big \},    
\ee
where $I=\int {d^2q^\perp \over {L_1 L_2}} $, $L_1=(q^\perp+(1-x) D^\perp)^2
+m^2 (1-x_1)^2$ and $L_2=
(q^\perp)^2+m^2 (1-x_2)^2$; $x_1={x-\xi\over 1-\xi}$, $
x_2={x+\xi\over 1+\xi}$. 
 
It is especially interesting to investigate (\ref{qnonflip})  in the forward 
limit. For simplicity, we consider the
massless case. 
We get, 
\be
F^{+q}_{++}(x,0,0) &=& {\mid \psi_1 \mid }^2 \delta (1-x)+\int 
d^2q^\perp \psi^*_{2+}(x, q^\perp) \psi_{2+}(x, q^\perp)\nonumber\\&& 
={\mid \psi_1 \mid }^2 \delta (1-x)+ {\mid \psi_1 \mid }^2 {\alpha\over 2
\pi} {1+x^2\over 1-x} log{\Lambda^2\over \mu^2},
\label{forward}
\ee
here $\mu$ is a scale, $\mu << \Lambda$.

The normalization in this case gives,
\be
{\mid \psi_1 \mid}^2=1-{\alpha\over {2 \pi}} 
\int_\epsilon^{1-\epsilon} dx {{1+x^2}\over {1-x}}log{\Lambda^2\over \mu^2}.
\label{nor}
\ee
Thus we have,
\be
F^{+q}_{++}(x,0,0) &=& \delta(1-x)+{\alpha\over 2\pi} log{\Lambda^2\over \mu^2}
\Big [ {1+x^2\over (1-x)_+} +{3 \over 2} \delta(1-x) \Big ].
\label{for}
\ee
Here the plus prescription is defined in the usual way. Now we know that in
the forward limit, $H_q(x,0,0)=q(x)$  which is the
(unpolarized) quark distribution of a given flavor in the proton. 
From (\ref{he}) we then get  $ F^{+q}_{++}(x,0,0)= q (x)$, so from
(\ref{for}) we get the splitting function for the leading order evolution of
the fermion distribution
\be
P_{qq}(x)={1+x^2\over 1-x}.
\ee  
Note that the $\epsilon$ dependence is no longer there in Eq. (\ref{for}).  
This result is also obtained in \cite{hari1} by calculating the structure
function of a quark dressed with a gluon (here one would also get  
the color factor $C_f$). In the nonforward case, one
obtains the splitting function for the LO evolution of the GPDs \cite{marc}.
Finally, it can be shown from (\ref{for}) that
\be
\int_0^1 dx H_q(x,0,0)= F_1(0)=1,
\ee
where $F_1(0)$ is the Dirac form factor at zero momentum transfer. 

Next we calculate the helicity flip matrix element
\be
F^{+q}_{+-}&=&\int {dy^-\over {8 \pi}} e^{{i\over 2} x
{\bar P}^+{y^-}}\langle P'
+ \mid \bar \psi (-{ y^-\over 2}) \gamma^+ \psi ({ y^-\over 2})
\mid P -\rangle.
\label{hflip}
\ee
   
Contribution to (\ref{hflip}) comes from the two-particle sector of the 
state. The mass cannot be neglected here. It can be written as
\be
F^{+q}_{+-} = \int d^2q^\perp \psi^*_{2+}({x-\xi\over 1-\xi},
q^\perp+(1-x) D^\perp) \psi_{2-}({x+\xi\over 1+\xi}, q^\perp).
\label{qflip}
\ee
Using Eq. (\ref{psi2}) we get,
\be
F^{+q}_{+-}&=& {e^2\over (2 \pi)^3} (im) \int {d^2 q^\perp\over L_1 L_2}
{1\over \sqrt {(1-x_1) (1-x_2)}} \Big [ (i q^1+q^2) [x_1 (1-x_2)^2-x_2
(1-x_1)^2] \nonumber\\&&~~+ (i D^1+D^2) (1-x) x_1 (1-x_2)^2 \Big ].
\ee

The $q^\perp$ integration can be performed either using the Feynman
parameter method or by using
\be
{1\over A^k}={1\over \Gamma(k)} \int_0^\infty d \beta \beta^{k-1} e^{-\beta
A}.
\ee  
Here we use the latter method and we get,
\be
\int d^2 q^\perp {(i q^1+q^2)\over L_1 L_2} &=& -i \pi\int_0^\infty d \sigma
\int_0^\infty d \beta {\beta\over (\sigma+\beta)^2} (1-x) D_V^\perp e^{-\sigma
{\beta \over \sigma+\beta} (1-x)^2 D_\perp^2} \nonumber\\&&~~~~~~~
e^{-m^2 [\sigma (1-x_2)^2+\beta (1-x_1)^2]}
\label{Eq}
\ee
where $D_V^\perp=D^1-i D^2$. We introduce the variables $\lambda$ 
and $y$ defined as,
\be
\lambda = \sigma+\beta,~~ \sigma= y \lambda,~~ \beta=(1-y) \lambda.
\ee
The above integral can be written as,
\be
\int d^2 q^\perp {(i q^1+q^2)\over L_1 L_2} &=& -i \pi \int_0^1 dy (1-y)
(1-x) D_V^\perp \int_0^\infty d \lambda e^{-\lambda y (1-y) (1-x)^2
D_\perp^2} e^{-\lambda m^2 [y (1-x_2)^2+(1-y) (1-x_1)^2]}\nonumber\\         
&=& -i \pi \int_0^1 dy {(1-y) (1-x) D_V^\perp \over {y (1-y) (1-x)^2
D_\perp^2 +m^2 [y (1-x_2)^2+(1-y) (1-x_1)^2]}}.
\ee 
The other integral can be done in a similar way :,
\be
\int d^2 q^\perp {i D_V^\perp\over L_1 L_2} &=& i \pi D_V^\perp
\nonumber\\&&~~\int_0^1 dy {1\over {y (1-y) (1-x)^2
D_\perp^2 + m^2 [y (1-x_2)^2+(1-y) (1-x_1)^2]}}.
\ee
Substituting in Eq. (\ref{Eq}) we obtain,
\be
F^{+q}_{+-}&=&{e^2\over (2 \pi)^3} \pi (i m) {(1-x)\over \sqrt {(1-x_1)
(1-x_2)}} (-i D^1 -D^2) \nonumber\\&&
\int_0^1 dy {[x_1 (1-x_2)^2-x_2 (1-x_1)^2](1-y)-x_1 (1-x_2)^2 \over 
{y (1-y) (1-x)^2
D_\perp^2 + m^2 [y (1-x_2)^2+(1-y) (1-x_1)^2]}}.
\ee
In the forward limit, $\xi=0$, $x_1=x_2=x, D^\perp=0$ and we obtain, using Eq.
(\ref{heflip})
\be
E_q(x,0)={\alpha\over \pi} \int_0^1 dy {x (1-x)^2 \over {y (1-x)^2+(1-y)
(1-x)^2}}= {\alpha\over \pi} x,
\ee
which gives the Schwinger value for the anomalous magnetic
moment of an electron in QED \cite{imp1}: 
\be
\int_0^1 E_q dx = F_2(0) = {\alpha\over 2 \pi}.
\ee
\section{Gauge Boson GPDs}
We calculate the helicity non-flip part of the gauge boson matrix element $
F^{+g}_{++}$ for the same state as before. This gives information on the
gauge boson GPDs $H_g$ and $E_g$. Contribution comes from the two-body wave
function, which has one fermion and one gauge boson as constituents. This
can be written as an overlap
\be
F^{+g}_{++} = \int d^2q^\perp \psi^*_{2+}({1-x\over 1-\xi},
q^\perp) \psi_{2+}({1-x\over 1+\xi}, q^\perp+(1-x) D^\perp).
\label{gnonflip}
\ee
The $q^\perp$ integral is divergent. The above can be calculated using
(\ref{psi2}) :
\be
F^{+g}_{++} &=& {e^2\over 2 (2 \pi)^3} 
  {1\over \sqrt {(1-x_1) (1-x_2)} } {\sqrt {x^2-\xi^2}\over x} 
\Big \{ 2 (1+ x_1
x_2) \nonumber\\&&~~~\Big [ \pi log[{\Lambda^2\over {(1-x)^2 D_\perp^2 
+m^2 (1-x_2)^2}}]-m^2 (1-x_1)^2 I \Big ]+ (1+x_1 x_2) \Big [ 
\pi log[{\Lambda^2\over m^2 (1-x_1)^2}] 
\nonumber\\&&~~~-\pi log[{\Lambda^2\over (1-x)^2 D_\perp^2 +m^2 (1-x_2)^2}] 
-\Big ( (1-x)^2
D_\perp^2 +m^2 [ (1-x_2)^2-(1-x_1)^2 ] \Big ) I \Big ]
\nonumber\\&&~~~~~~~~~+ 2 m^2 (1-x_2)^2
(1-x_1)^2 I \Big \},    
\ee
where $I=\int {d^2q^\perp \over {L_1 L_2}} $, $L_2=(q^\perp+(1-x) D^\perp)^2
+m^2 (1-x_2)^2$ and $L_1=
(q^\perp)^2+m^2 (1-x_1)^2$; $x_1={1-x\over 1-\xi}$, $
x_2={1-x\over 1+\xi}$. 

Like the quark case, it is again interesting to look at the forward limit of
the above expression. We get
\be
F^{+g}_{++} (x,0,0) = {\alpha\over 2 \pi} log{\Lambda^2\over \mu^2}
{1+(1-x)^2\over x}.
\ee
Here we have neglected the electron mass for simplicity. $F^{+g}_{++}
(x,0,0)$ gives the (unpolarized) gluon distribution in the nucleon and  
the above expression gives the splitting function \cite{hari1}
\be
P_{gq}(x)={1+(1-x)^2\over x}.
\ee   
In the off-forward case, the splitting functions can be found 
using the same approach \cite{marc}.

We now calculate the helicity flip gauge boson GPD $E_g$ given in
(\ref{gpdg}), for the same state. 
The matrix element is given by,
\be
F^{+g}_{+-}&=&{1\over {8 \pi x {\bar P}^+}}\int dy^- 
e^{{i\over 2}{\bar P}^+ { y^-} x}\langle P'
+ \mid F^{+ \alpha}(-{ y^-\over 2}) {F^+}_\alpha ({ y^-\over 2})
\mid P - \rangle. 
\ee
Contribution comes from the two particle sector. As in the fermion case, the
mass cannot be neglected here. This can be written as,
\be
F^{+g}_{+-} = \int d^2q^\perp \psi^*_{2+}({1-x\over 1-\xi},
q^\perp) \psi_{2-}({1-x\over 1+\xi}, q^\perp+(1-x) D^\perp).
\label{gflip}
\ee
Using Eq. (\ref{psi2}) we get,
\be
F^{+g}_{+-}&=&{e^2\over (2 \pi)^3} (i m) {\sqrt {x^2-\xi^2}\over x} \Big [
\int d^2 q^\perp {i q^\perp_V\over L_1 L_2 } [x_1 (1-x_2)^2-x_2 (1-x_1)^2]
\nonumber\\&& -(1-x) x_2 (1-x_1)^2 i D_V^\perp \int {d^2 q^\perp\over L_1
L_2}.  
\ee
The $q^\perp$ integration can be performed in a similar way as for the
fermions and we get,
\be
F^{+g}_{+-}&=& {\alpha\over 2 \pi} m {\sqrt {1-\xi^2}\over x} D_V^\perp
(1-x) \nonumber\\&& \int_0^1 dy {[x_1 (1-x_2)^2-x_2 (1-x_1)^2] (1-y) 
+x_2 (1-x_1)^2\over y
(1-y) (1-x)^2 D_\perp^2 +m^2 [y (1-x_1)^2+(1-y) (1-x_2)^2]}.
\ee  
In the forward limit, $x_1=x_2=1-x$ and we get, in agreement with
\cite{imp1}, 
\be
E_g (1-x,0)= -{\alpha\over \pi} {(1-x)^2\over x},
\ee
here $E_g$ is given by Eq. (\ref{heflip}). Note that $x$ in the forward case
is the momentum fraction of the gauge boson.
The second moment of $E_{q,g}(x,0), \int dx x E_{q,g}(x,0)$ gives in units
of ${1\over 2m}$ by how much the transverse center of momentum of the parton
$q,g$ is shifted away from the origin in the transversely polarized state.
When summed over all partons, the transverse center of momentum would still
be at the origin. Indeed it is easy to check for a dressed electron
\cite{imp1}
\be
\int_0^1 dx x E_q(x,0)+\int_0^1 dx (1-x) E_g (x,0)=0,
\ee
which is due to the fact that the anomalous gravitomagnetic moment of the
electron has to vanish \cite{bur1}. Note that in the second term, 
$(1-x)$ is the momentum fraction of the gauge boson.
\section{GPDs in the impact parameter Space}
Fourier transform of the GPDs with respect to the transverse momentum
transfer $\Delta^\perp$ brings them to the impact parameter space. When the 
longitudinal momentum transfer $\xi=0$, this gives the density of partons
with longitudinal momentum fraction $x$ and transverse distance $b$ from the
center of the proton. For non-zero $\xi$, the Fourier transforms are defined
as \cite{diehl},
\be
I_{++}^{+ q,g} (x,\xi,b^\perp) &=& \int {d^2 D^\perp\over (2 \pi)^2} e^{-i
D^\perp \cdot b^\perp} F_{++}^{+ q,g}(x,\xi, D^\perp)\nonumber\\&&
={1\over 4 \pi} \int_0^\infty d (D^\perp)^2 J_0 (\mid D \mid  \mid b \mid )
(H_{q,g}- {\xi^2\over 1-\xi^2} E_{q,g})
\label{ftnonflip}
\ee
\be
I_{+-}^{+ q,g} (x,\xi,b^\perp) &=& \int {d^2 D^\perp\over (2 \pi)^2} e^{-i
D^\perp \cdot b^\perp} F_{+-}^{+ q,g}(x,\xi, D^\perp)\nonumber\\&&
={1\over 4 \pi} {b^2-i b^1\over \mid b^\perp \mid} 
\int_0^\infty d (D^\perp)^2 J_1 (\mid D \mid  \mid b \mid ) {\mid D \mid
\over 2 m} E_{q,g};
\label{ftflip}
\ee
where $J_0$ and $J_1$ are Bessel functions and $b^\perp$ is called the
impact parameter. In order to avoid infinities in
the intermediate steps, we take a wave packet state 
\be
\int {d^2 p^\perp \over 16 \pi^3} \phi(p^\perp) \mid p^+, 
p^\perp,\lambda \rangle,
\ee
which have a definite plus momentum. 
Following \cite{diehl} we take a Gaussian wavepacket,
\be
\phi(p^\perp)= G(p^\perp,\sigma^2)
\ee

where 
\be
G(p^\perp,\sigma^2)= e^{-(p^\perp)^2\over 2 \sigma^2}.
\ee
$\sigma$ gives the width of the wave packet. It is the accuracy
to which one can localize information in the impact parameter space
\cite{diehl}. 
The states centered around $b_0$ with an accuracy $\sigma$  
are normalized as,
\be
\langle p'^+, b'^\perp, \lambda' \mid p^+, b^\perp, \lambda \rangle = {1\over
16 \pi^2 \sigma^2 }{1\over p^+} G(b'^\perp-b^\perp, 2 \sigma^2)
\delta(p^+-p'^+).
\ee
Fourier transform of the matrix elements in Eq. (\ref{ftnonflip}) and
(\ref{ftflip}) with the Gaussian wave packet probe partons in the nucleon at
transverse position $b$ but when the initial and final state protons are
centered around $0$ but shifted from each other by an amount of the order of
$\xi \mid b^\perp \mid $. In our case, they probe a bare electron or a
photon in an electron dressed with a photon.

It is interesting to look at the qualitative behavior of the helicity-flip
GPDs $E_q$ and $E_g$ in the impact parameter space. The contribution in this
case comes purely from the two-body sector, that is it involves the wave
function $\psi_2$ of the relativistic spin-${1/2}$ system. The overlap is
given in terms of the light cone wave functions whose orbital angular
momentum differ by $\Delta L_z= \pm 1$ \cite{brod1}.  The scale
dependence, as mentioned before, is suppressed here, unlike $H_q$ and $H_g$.  
We use the notation 
\be
{\mathcal E}_{q,g} (x, \xi, b^\perp)={1\over 4 \pi} 
\int_0^\infty d (D^\perp)^2 J_0 (\mid D \mid  \mid b \mid ) E_{q,g};
\ee
\be
{\mathcal E}^1_{q,g} (x, \xi, b^\perp)
={1\over 4 \pi}  
\int_0^\infty d (D^\perp)^2 J_1 (\mid D \mid  \mid b \mid ) {\mid D \mid
\over 2 m} E_{q,g}
\ee
which contribute to Eq. (\ref{ftnonflip}) and (\ref{ftflip}) respectively. 
Fig. 1(a) shows ${\mathcal E}_q$ vs $ b^\perp $ for fixed $\xi=0.1$ 
and three different values of $x> \xi$. We have plotted for positive $b^\perp$.
The functions are symmetrical in $\mid b^\perp \mid $.  
For all numerical studies, we have 
taken a Gaussian wave packet. ${\mathcal E}_q$  
is positive and has a maximum for $\mid
b^\perp \mid =0$ and decreases smoothly with increasing $\mid b^\perp \mid
$. For fixed $\mid b^\perp\mid$, ${\mathcal E}_q$ is higher in magnitude for
higher $x$. The qualitative behaviour for nonzero
$\xi$ is the same as $\xi=0$. We have taken the normalization to be
${\alpha\over 2 \pi}=1$ and $m=0.5$ MeV. Fig. 1 (b)
shows ${\mathcal E}_q$ vs $\xi$ for fixed $b^\perp=0.1 {\mathrm{MeV}}^{-1}$ and three different
values of $x>\xi$. ${\mathcal E}_q$ is again a smooth function of $\xi$ and
increases as $\xi$ increases for given $b^\perp$ and $x$. For fixed $\xi$, 
${\mathcal E}_q$ increases as $x$ increases. 
We have plotted ${\mathcal E}^1_q
$ in Fig. 1 (c) as a function of $b^\perp$ for $\xi=0.1$ and three different
values of $x$. The rise near $b^\perp=0$ is much more sharp here than in Fig. 1(a).

In Fig. 2 (a) we have shown ${\mathcal E}_g (x, \xi, b^\perp) $ 
vs $b^\perp$ for fixed
$\xi=0.1$ and for three different values of $x$. ${\mathcal E}_g$ is
negative for positive $b^\perp$. It has a negative maximum at $b^\perp=0$
and smoothly decreases in magnitude as $b^\perp$ increases. Again, the
qualitative behaviour is the same as $\xi=0$. For a given $b^\perp$,  
${\mathcal E}_g$ decreases in magnitude for increasing $x$. 
Fig. 2 (b) shows  ${\mathcal E}_g$ as a function
of $x$ for fixed $b^\perp=0.1 {\mathrm{MeV}}^{-1}$  and three different values of $\xi<x$. 
 ${\mathcal E}_g$ vanishes at $x=1$. For $\xi=0$, it also vanishes at $x=0$. For
$\xi>0$,  ${\mathcal E}_g$ increases in magnitude as $\xi$ increases. The
curves cannot be continued for $x<\xi$ as there will be contribution from
the higher Fock components in this region. Fig. 2 (c) shows ${\mathcal E}^1_g$ vs $b^\perp$
for $\xi=0.1$ and three different values of $x$. The qualitative behaviour is
the same as in (a), again the rise near $b^\perp=0$ is much sharp like the
fermion case. For large $b^\perp$, ${\mathcal E}^1_g$ behaves in the same
way independent of $x$.  For completeness, in Fig. 3 (a) and 3
(b) we show $H_q(x,0,t)$ and $H_g( x,0, t)$ in the impact parameter space,
${\mathcal H}_q (x, b^\perp)$ and ${\mathcal H}_
g (x, b^\perp)$ respectively for a definite scale $\Lambda=5$ GeV. The width of the 
Gaussian is $\sigma=0.1$. Both of them are smooth functions of
$b^\perp$, increases as $\mid b^\perp \mid $ decreases. We have omitted 
the very small $b^\perp$ region in order to show the resolution of the 
different curves at higher $b^\perp$. For a given $b^\perp$, ${\mathcal H}_
q (x, b^\perp) $ increases as $x$ becomes closer to $1$ and at $x
\rightarrow 1$ it becomes a delta function \cite{imp1} which can be seen
analytically.
${\mathcal H}_g (x, b^\perp)$ on the other hand, decreases in magnitude for
a given $b^\perp$ as $x$ goes closer to $1$. 

\section{Inequalities}

GPDs in the impact parameter space obey certain inequalities, which impose
severe constraints on phenomenological models of GPDs. The  most
general forms of these inequalities were derived in \cite{pob} from
positivity constraints. The spin flip GPDs ${\mathcal E}_{q,g} (x, b^\perp)$
defined as the Fourier transform of  ${ E}_{q,g} (x, b^\perp)$ for $\xi=0$
obey two inequalities given in \cite{ineq}, both of them can be shown to
hold for a dressed electron state. For non-zero $\xi$, a general inequality
can be derived \cite{diehl},
\be
(1-\xi^2)^3 {\mid I_{\lambda' \lambda}^{q,g} (x, \xi, b^\perp) \mid }^2
\le  I_{++}^{q,g} ({x-\xi\over 1-\xi}, 0, {b^\perp\over 1-\xi})
I_{++}^{q,g} ({x+\xi\over 1-\xi}, 0, {b^\perp\over 1+\xi}),
\label{ineq}
\ee 
for $\xi \le x \le 1$ and for any combinations of helicities $\lambda',
\lambda$. For $\lambda'\ne \lambda $, the inequality is easy to prove, as
there are large logarithmic contribution to $ I_{++}^{q,g} $ from
the scale dependent part. For $\lambda'=\lambda $, the inequality is
non-trivial and can be verified numerically. Fig. 4 shows the {\it lhs} and
{\it rhs} of Eq. (\ref{ineq}) vs $b^\perp$ for the fermions for 
two different choices of the scale      
$\Lambda=7$ GeV and $\Lambda=5$ GeV. The scale dependence comes
entirely from $H_q$. In the plot $x_1={x-\xi\over 1-\xi}, x_2={x+\xi\over 1-\xi},
b_1={b^\perp\over 1-\xi}$ and $b_2={b^\perp\over 1+\xi}$. We have taken
$\xi=0.1 $ and $x=0.5$ and $\sigma=0.1$.      
\section{Summary and discussions}
In this work, we have investigated the GPDs for an effective spin-$1/2$
composite relativistic system, namely for an electron dressed with a photon
in QED.  It is known \cite{brod1} that the light-cone wave function of 
this two-body  state gives a template
for the effective quark spin-one diquark structure of the proton light cone
wave function, which provides the phenomenological relevance of our
study. The GPDs are expressed as overlaps of the light-cone wave
functions, which in this case are known order by order in perturbation
theory. We keep both the skewedness $\xi$ and the transverse momentum
transfer $\Delta^\perp$ non-zero, which is relevant for deeply virtual
Compton scattering experiments to probe the GPDs. Fourier transform with
respect to the transverse momentum transfer brings the GPDs to the impact    
parameter space. We showed that the GPDs for the effective state obey the
necessary inequalities and investigated the qualitative behaviour of the 
fermion and the gauge boson GPDs in the impact parameter space. 
\section{acknowledgment}

We thank M. Diehl and P. Pobylitsa for fruitful discussions and
correspondence. 
The work of DC was partially supported by the Department of Energy under
Grant No. DE-FG02-97ER-41029 and the work of AM has been supported by
FOM, Netherlands.  
   

\newpage
\begin{figure}
\centering
\psfrag{A}{$b_\perp (MeV^{-1})$}
\psfrag{B}{${\mathcal E}_q(x,\xi,b_\perp)$}
\psfrag{C}{$\xi$}
\includegraphics[width=6.7cm,clip]{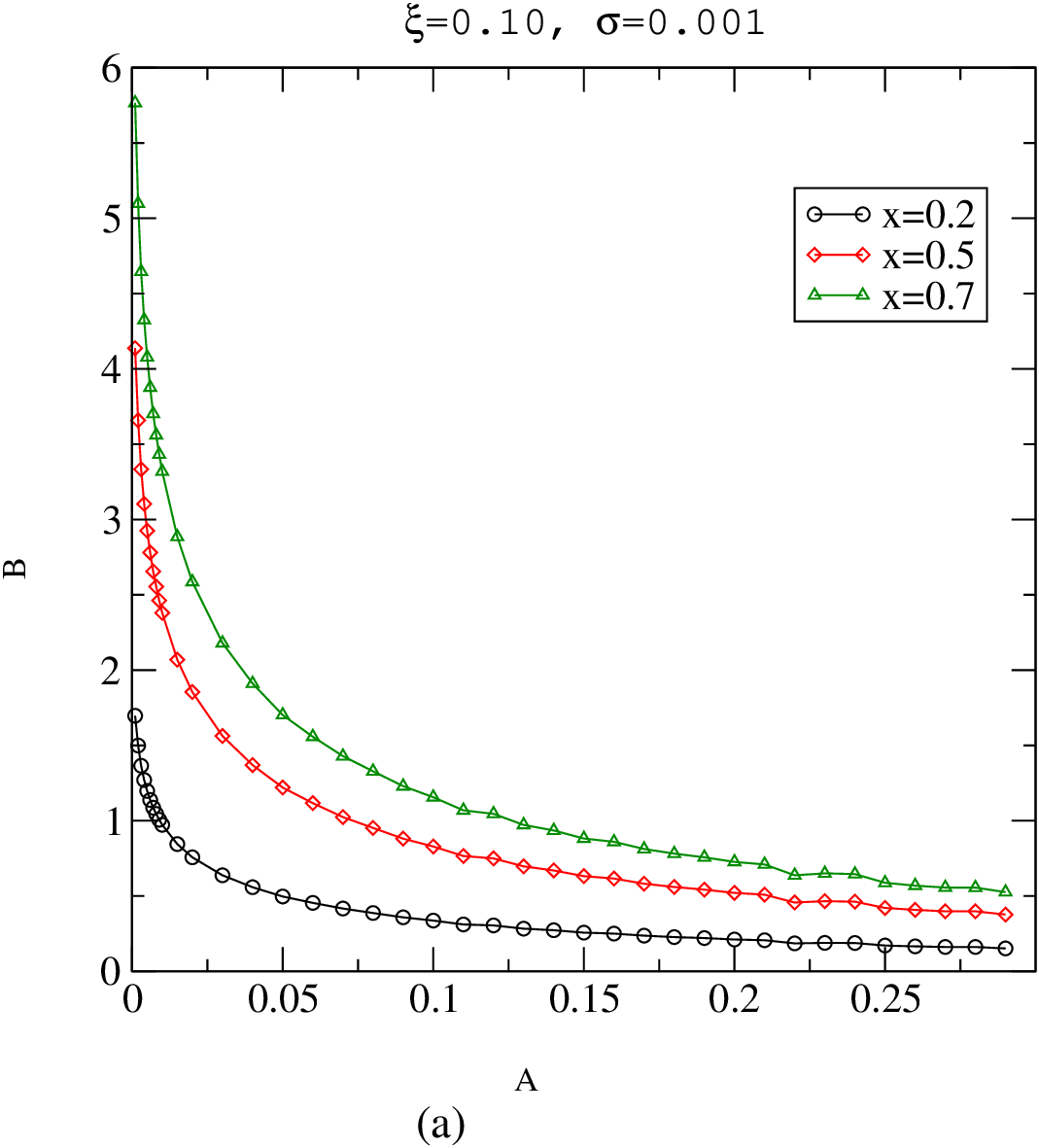}%
\hspace{0.2cm}
\includegraphics[width=7cm,clip]{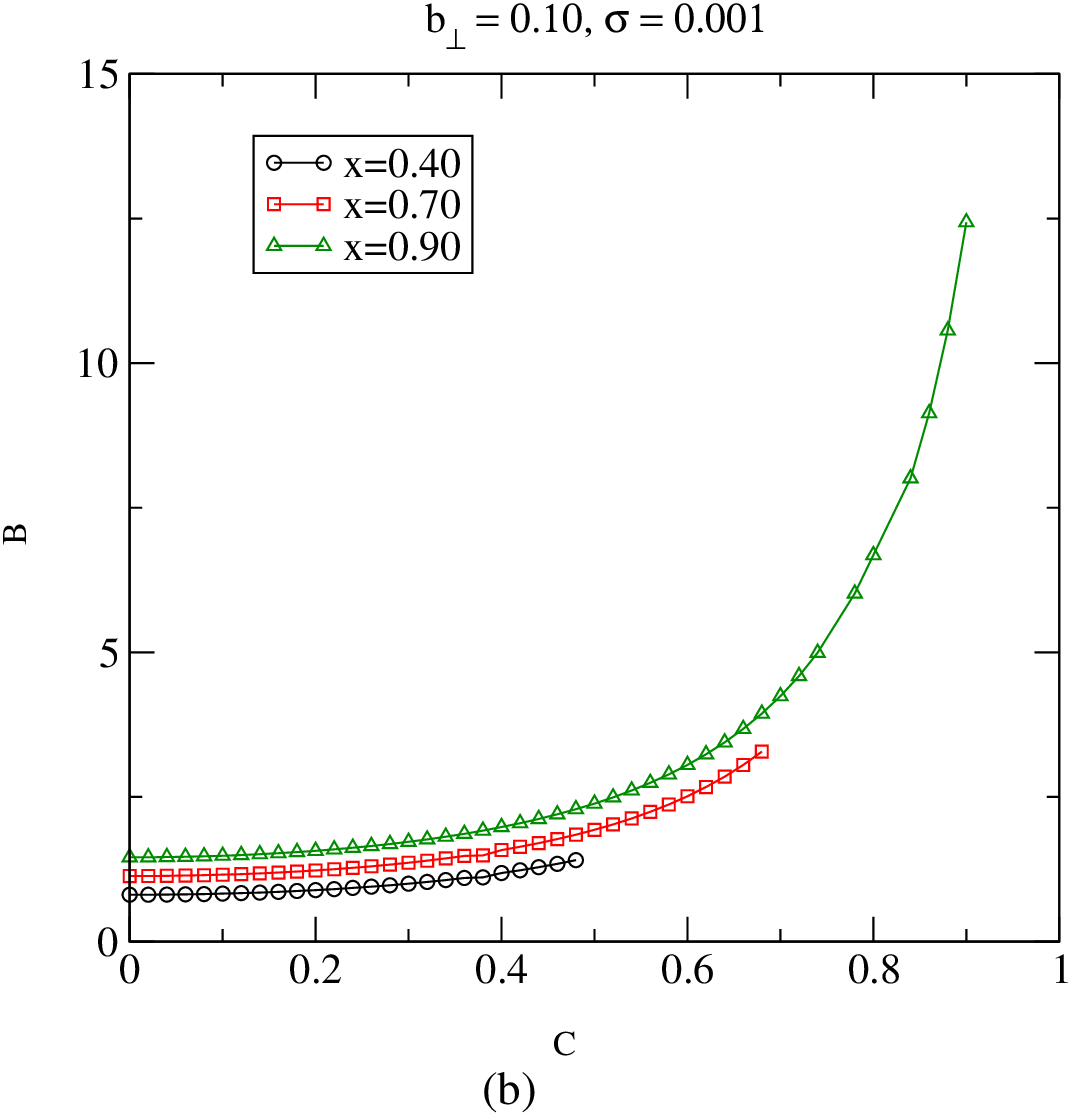}
\end{figure}

\vspace{.4cm}
\begin{figure}
\centering
\begin{minipage}[c]{0.5\textwidth}
\centering
\psfrag{A}{$b_\perp (MeV^{-1})$}
\psfrag{B}{${\mathcal E}_q^1(x,\xi,b_\perp)$}
\includegraphics[width=7cm,clip]{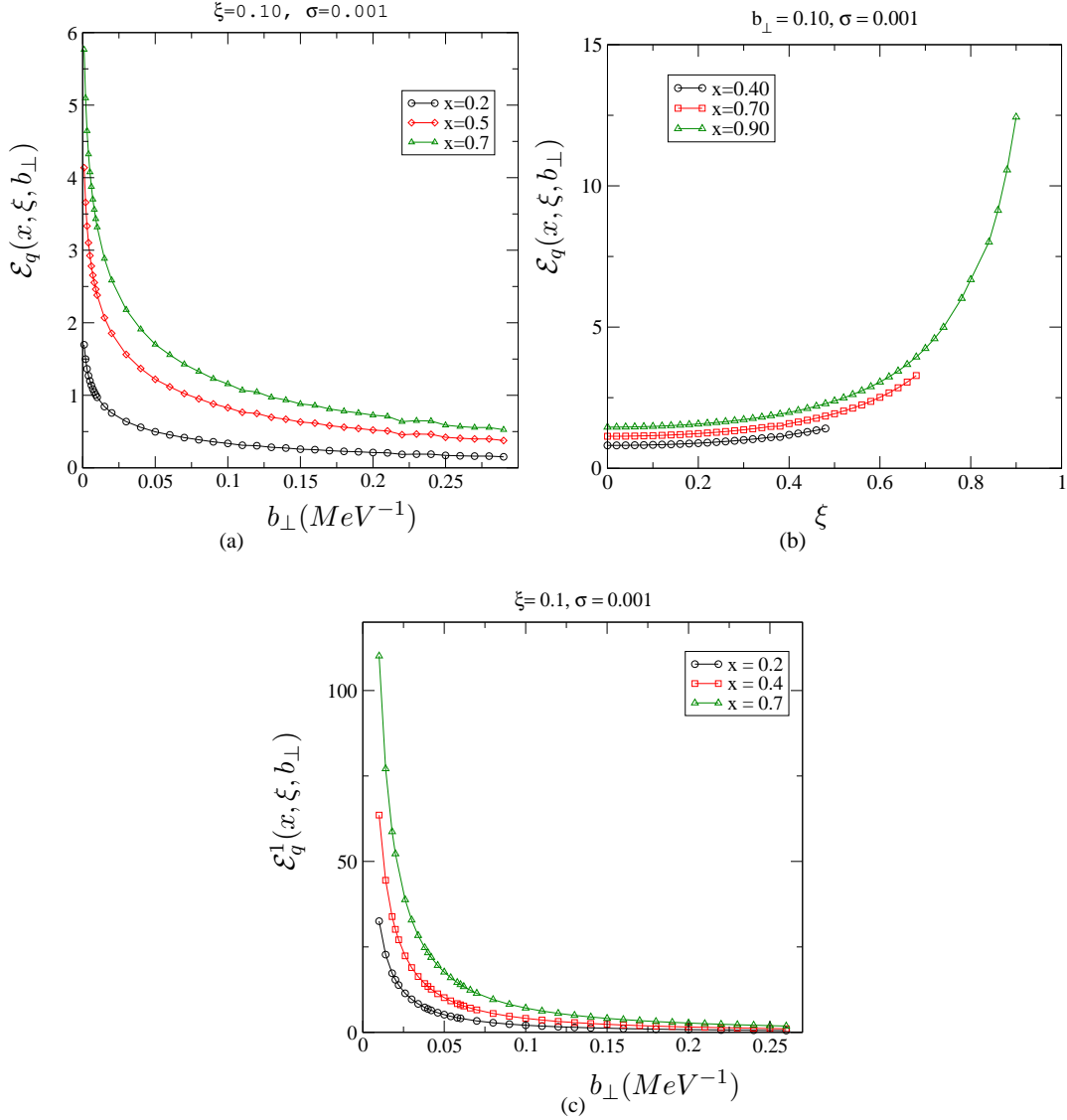}
\end{minipage}%

\caption{\label{fig1}(a) ${\mathcal E}_q$ vs $ b^\perp $ for 
$\xi=0.1$ , (b) ${\mathcal E}_q$ vs $\xi$ 
for $b^\perp=0.1 {\mathrm{MeV}^{-1}}$, (c) 
${\mathcal E}^1_q$  vs $b^\perp$ for $\xi=0.1$ and three different values of
$x$. We have taken $\sigma=0.001$.}   
\end{figure}

\vspace{4cm}
\newpage
\begin{figure}
\centering
\psfrag{A}{$b_\perp (MeV^{-1})$}
\psfrag{B}{${\mathcal E}_g(x,\xi,b_\perp)$}
\psfrag{C}{$x$}
\includegraphics[width=7cm,clip]{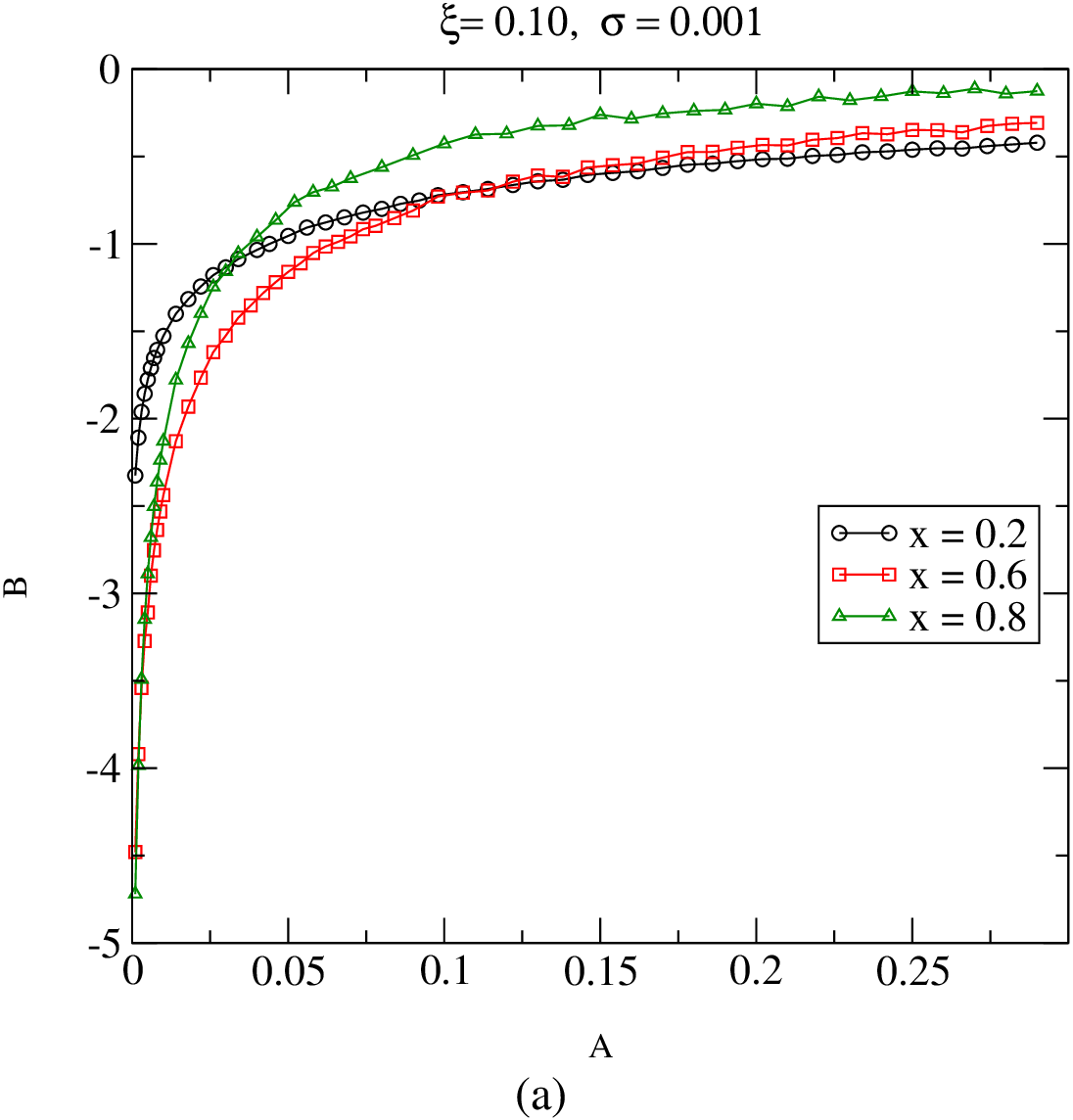}%
\hspace{0.2cm}%
\includegraphics[width=7cm,clip]{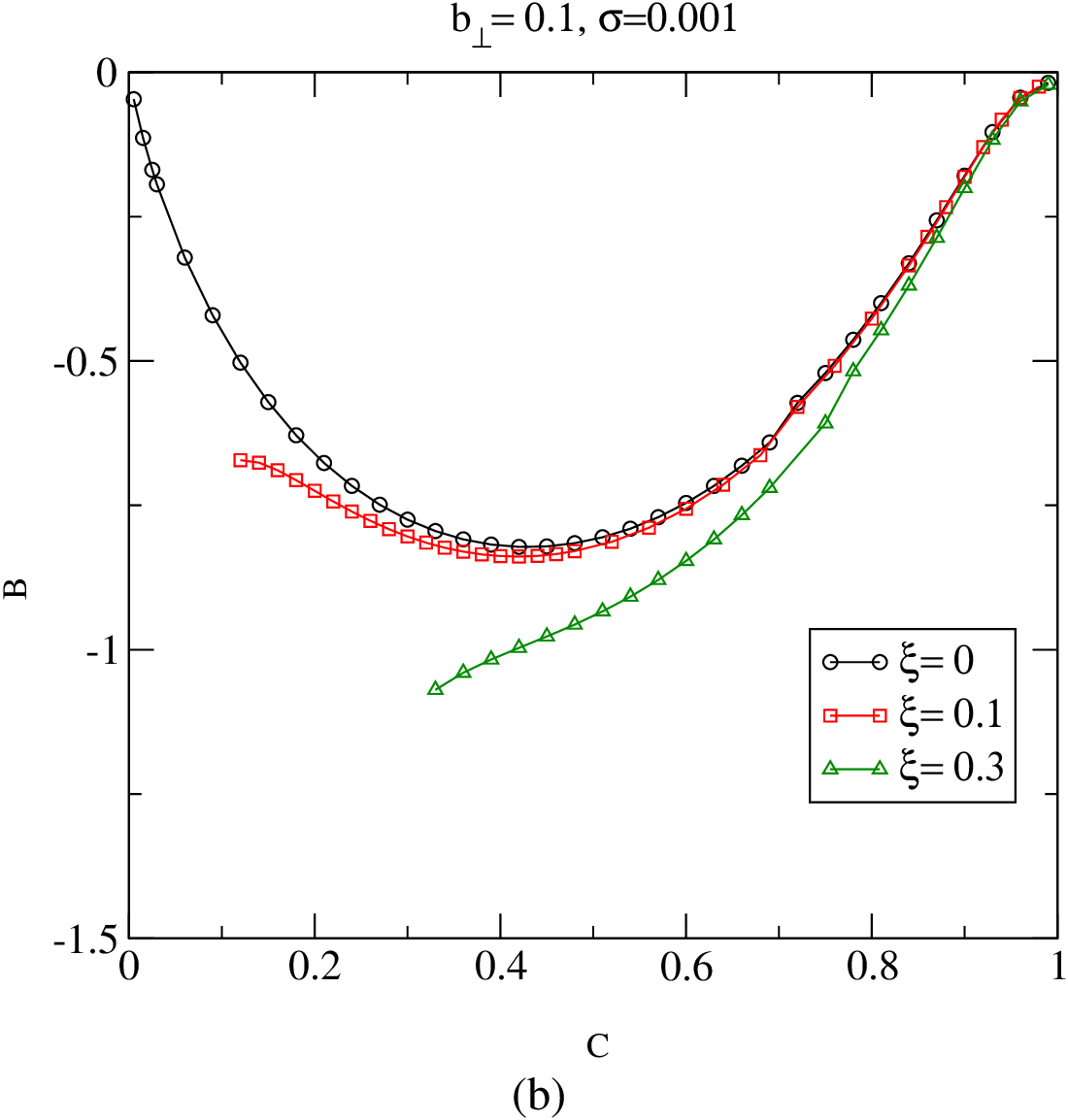}
\end{figure}

\begin{figure}
\centering
\begin{minipage}[c]{0.5\textwidth}
\centering
\psfrag{A}{$b_\perp (MeV^{-1})$}
\psfrag{B}{${\mathcal E}_g^1(x,\xi,b_\perp)$}
\includegraphics[width=7cm,clip]{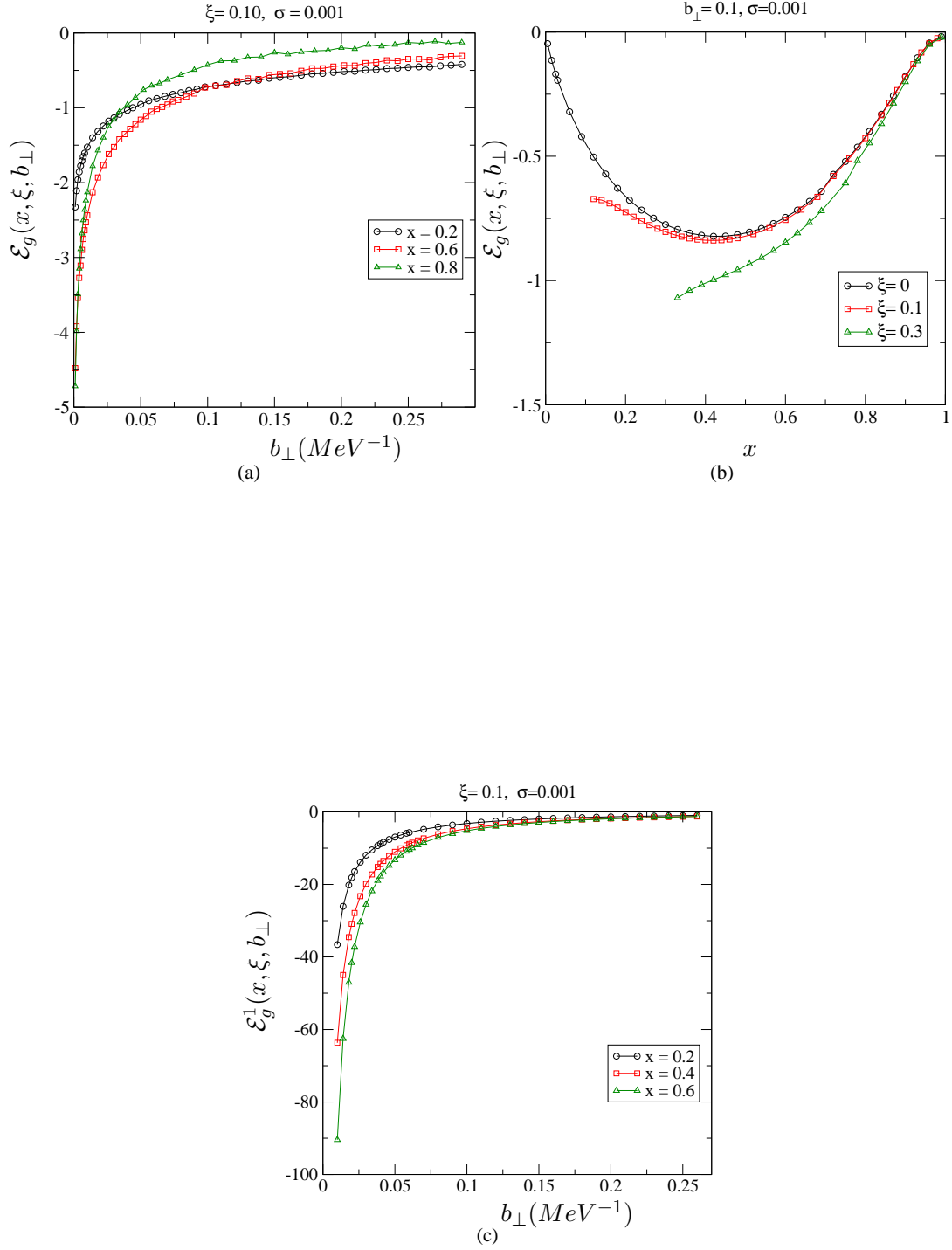}
\end{minipage}%
\caption{\label{fig2}(a) ${\mathcal E}_g$ vs $ b^\perp $ for 
$\xi=0.1$ and three different values of $x$. (b) 
${\mathcal E}_g$ vs $x$ 
for $b^\perp=0.1 {\mathrm{MeV}}^{-1}$ and three different values of $\xi$. (c) 
${\mathcal E}^1_g$ vs $b^\perp$ 
with $\xi=0.1$ and three different values of $x$. We have taken
$\sigma=0.001$. }   
\end{figure}

\vspace{4cm}
\newpage
\begin{figure}
\centering
\begin{minipage}[c]{0.9\textwidth}
\centering
\psfrag{A}{$b_\perp~(MeV^{-1})$}
\psfrag{B}{${\mathcal H}_q(x, 0, b_\perp)$}
\psfrag{C}{${\mathcal H}_g(x, 0, b_\perp)$}
\includegraphics[width=7cm,clip]{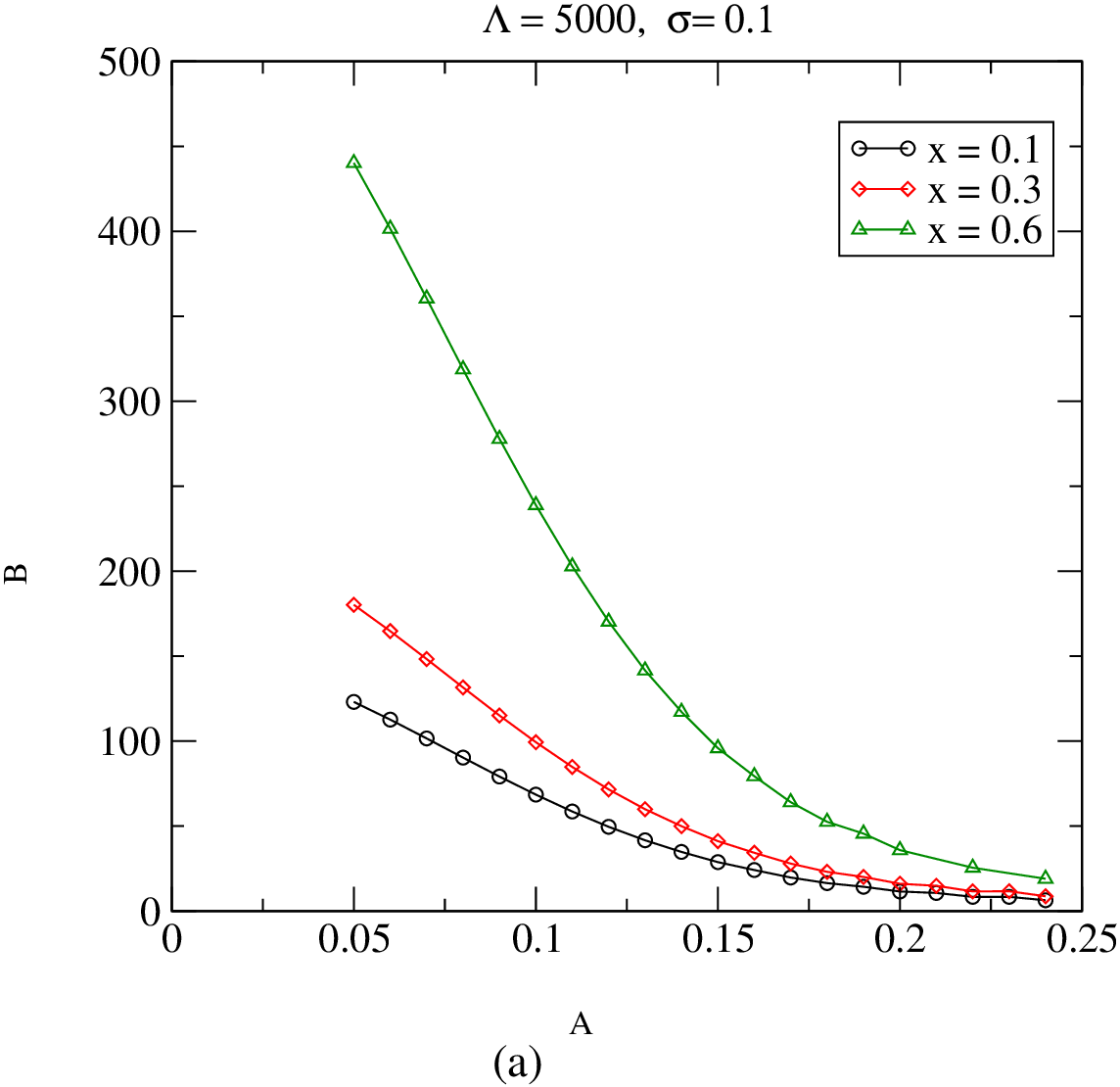}%
\hspace{0.2cm}%
\includegraphics[width=7cm,clip]{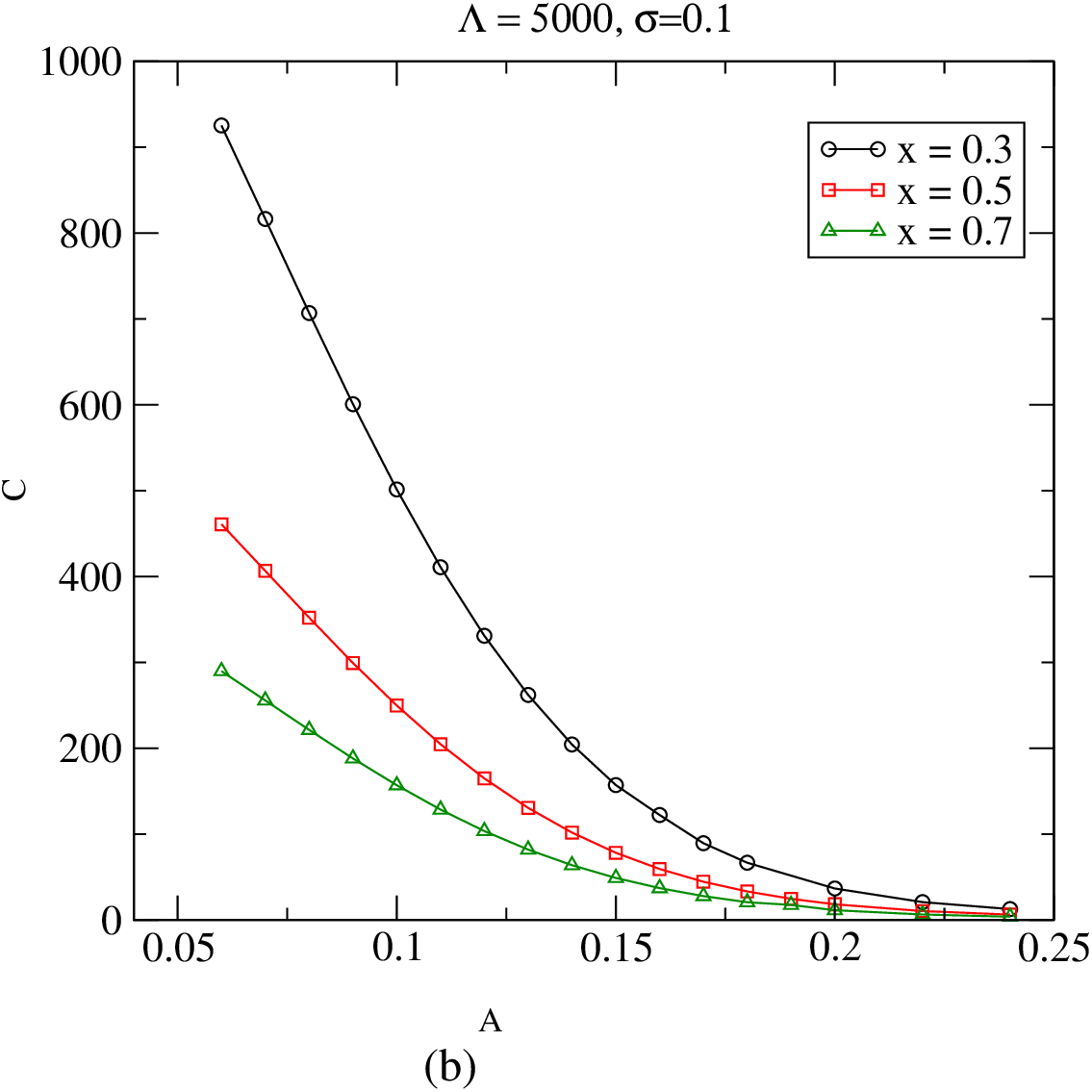}
\end{minipage}%
\caption{\label{fig3} (a) ${\mathcal H}_q (x, b^\perp)$ vs $b^\perp$, (b) 
${\mathcal H}_g (x, b^\perp)$ vs. $b^\perp$ for three different values of $x$.
$\Lambda$ is given in MeV.} 
\end{figure}

\vspace{0.4cm}

\begin{figure}
\centering
\begin{minipage}[l]{0.5\textwidth}
\centering
\psfrag{A}{$b_\perp (MeV^{-1})$}
\includegraphics[width=8cm]{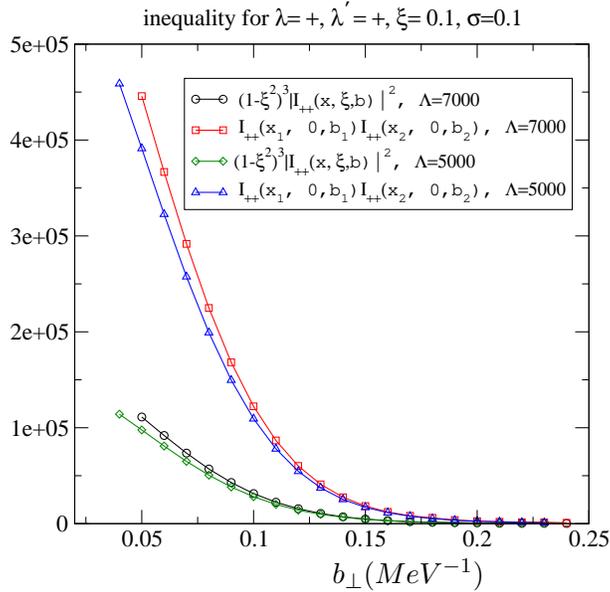}
\end{minipage}%
\caption{\label{fig4} Inequality for $I_{++}$. $\Lambda$ is given in MeV. Here
$x_1={x-\xi\over 1-\xi}$, $b_1={b^\perp \over 1-\xi}$, $x_2={x+\xi\over 1+\xi}$, 
$b_2={b^\perp \over 1+\xi}$.} 
\end{figure}

\end{document}